\documentstyle[aps,prl,graphicx,multicol]{revtex} 
\begin{document}
\draft
\title{Interacting Monomer-Dimer Model with 
Infinitely Many Absorbing States}
\author{WonMuk Hwang}
\address{Department of Physics, 
Boston University, Boston, MA 02215.}
\author{Hyunggyu Park}
\address{Department of Physics, Inha University,  Inchon 402-751, Korea}

\date{\today}
\maketitle            
\begin{abstract}
We study a modified version of the interacting monomer-dimer (IMD)
model that has infinitely many absorbing (IMA) states. Unlike
all other previously studied models with IMA states, 
the absorbing states can be divided into two equivalent groups which
are dynamically separated infinitely far apart.
Monte Carlo simulations show that
this model belongs to the directed Ising
universality class like the ordinary IMD model with two equivalent
absorbing states. This model is the first model with IMA states
which does not belong to the directed percolation (DP) universality class. 
The DP universality class can be restored in two ways, 
i.e., by connecting the two equivalent groups dynamically or
by introducing a symmetry-breaking field between the two groups.
\end{abstract}


\begin{multicols}{2}
\narrowtext

A wide variety of nonequilibrium systems 
with a single trapped (absorbing)
state display a continuous phase transition from
an active phase into an absorbing phase, which 
belongs to the directed percolation (DP) universality 
class~\cite{Marro-96,Grass-Torre-79,Cardy-Sugar-80,Grin-Lai-Browne-89}.
Recently, systems with multiple absorbing states have been 
investigated extensively. 
The interacting monomer-dimer(IMD) model 
introduced by one of us~\cite{Park-94}
is one of many models that have two equivalent absorbing 
states~\cite{Grass-84,Menyhard-94,Bassler-96,Hinrichsen-97}.
These models belong to a different universality class from 
DP. By the analogy to the equilibrium Ising model that has two equivalent 
ground states, this new class is called as the directed Ising (DI) 
universality class~\cite{Park-98}. 
When the (Ising) symmetry between the absorbing 
states is broken in the sense that one of the absorbing states is
probabilistically preferable, the system goes back to the DP class\cite{Park-95d}. 
Hence, the symmetry between the absorbing states is the
key factor in determining the universality class of models with several
absorbing states. Unfortunately, no models with higher symmetries than the Ising
symmetry (like the three-state Potts symmetry) 
are found to have a stable absorbing phase as yet. 

In contrast, systems with infinitely many absorbing (IMA) states
are far less understood. All IMA systems studied so far belong to the DP
universality class\cite{Jensen-93c,Mendes-94}. The number of absorbing states
of these IMA systems grows exponentially with system size but there is
no clear-cut symmetry among absorbing states. 
Recently, it was argued that the IMA models should
belong to the DP class unless they possess any extra symmetry 
among absorbing states~\cite{Grass-95,Munoz-96}. 
However, no IMA model with an additional
symmetry has been studied up to date and the role of the symmetry
in the IMA systems is still unclear. 

In this Letter, we introduce an IMA model with the Ising symmetry
between two groups of absorbing states. 
These two groups of absorbing states are equivalent and
dynamically separated infinitely far apart.
In other words, 
an absorbing state in one group can not be reached
from any absorbing state in the other group by a
finite number of successive local changes. 
There is no infinite dynamic barrier among absorbing states 
inside each group. This dynamic barrier is similar to the free energy
barrier between ground states of equilibrium systems that
exhibit spontaneous symmetry breaking in the ordered phase.
Our numerical simulations show that this model belongs to the DI universality
class. Furthermore, we find that this model crosses over to the DP class
by allowing that the two absorbing groups are connected dynamically 
and/or by introducing a symmetry-breaking field to make one absorbing group
probabilistically preferable to the other.

Our model is a modified version of the ordinary IMD model that
we call the IMA-IMD model. Dynamic rules of the IMA-IMD model
are almost the same as those of the IMD model with infinitely
strong repulsion between the same species in one dimension~\cite{Park-94}. 
A monomer ($A$) cannot adsorb at a nearest-neighbor site of an already-occupied
monomer (restricted vacancy) but adsorbs at a free vacant site with no adjacent 
monomer-occupied sites. Similarly, a dimer ($B_2$) cannot adsorb at a 
pair of restricted vacancies ($B$ in nearest-neighbor sites) but adsorbs at a 
pair of free vacancies. There are no nearest-neighbor restrictions in 
adsorbing particles of different species. Only the adsorption-limited reactions 
are considered. Adsorbed dimers dissociate and a nearest neighbor 
of the adsorbed $A$ and $B$ particles reacts,
forms the $AB$ product, and desorbs the catalytic surface immediately.
Differentiation between the IMA-IMD model and the IMD model
comes in when there is an $A$ adsorption attempt at a vacant site between 
an adsorbed $A$ and an adsorbed $B$. In the IMD model,
we allow the $A$ to adsorb and react with the neighboring $B$, so
there are two equivalent absorbing states comprised of only monomers
at alternating sites, i.e., $(A0A0\cdots)$ and $(0A0A\cdots)$ 
where ``$0$'' represents a vacancy. In the IMA-IMD model, this process is 
disallowed. Then, any configuration can be an absorbing state if there are
no nearest neighbor pair of vacancies and no single vacany between two $B$ 
particles, e.g., ($\cdots B0A0BB0A0A\cdots$). To impose the Ising symmetry between the
absorbing states, we introduce the probability $s$ of spontaneous
desorption of a nearest neighbor pair of adsorbed $B$ particles. 
At finite $s$, an absorbing configuration cannot have 
this $BB$ pair. Hence only those configurations that
have particles at alternating sites and no two $B$'s at consecutive
alternating  sites become absorbing states, e.g., ($A0A0B0A0\cdots$) and
($0A0A0B0A\cdots$). The absorbing states are divided into 
two groups with particles occupied at odd- and even-numbered sites
(O group and E group).
The number of absorbing states in each group grows exponentially
with system size and there is a one-to-one mapping between absorbing states
in two groups. It is clear that one can not reach from 
an absorbing state in one group to an absorbing state in the other group
by a finite number of successive local changes. Any interface 
(active region) between two absorbing states in the different groups 
never disappears by itself in a finite amount of time, so there is an
infinite dynamic barrier between the two groups. These interfaces
annihilate pairwise only. 

The order parameter characterizing the absorbing phase transition 
is the density of active sites or kinks (domain walls). 
In the IMD model, the dimer density served well as the order parameter,
but it cannot do in this model. We use the kink density as the order
parameter. Kinks are defined such that 
all absorbing configurations have no kinks
but any local change of the absorbing configurations should produce
kinks. In this model, one should examine, at least, three adjacent sites
to check the existence of kinks. 
There are 13 possible configurations for three adjacent sites.
We assign a kink to eight different configurations; $000$, 
$00A$, $A00$, $B00$, $00B$, $B0B$, $BB0$, and $0BB$. 
Five others, $A0A$, $A0B$, $B0A$, $0A0$, and $0B0$, do not
have a kink. In this kink representation, there is no mod(2)
conservation of the total number of kinks.

Three independent critical exponents characterize the critical behavior
near the absorbing transition: 
the order parameter exponent $\beta$, correlation length exponent $\nu_\bot$, and 
relaxation time exponent $\nu_\|$~\cite{Grass-Torre-79}. 
Elementary scaling theory combined with the finite size scaling 
theory~\cite{Aukrust-90} predicts that the kink density $\rho(p_{c},L)$ at
criticality in the (quasi)steady state scales with system size $L$ as 
\begin{equation}
\rho(p_{c},L)\sim L^{-\beta/\nu_{\perp}}.\label{rhoL}
\end{equation}
One can also expect the short time behavior  of the kink density
as $\rho(p_{c},t)\sim t^{-\beta/\nu_{\|}}$ and the 
characteristic time scales with system size as
$\tau(p_{c},L)\sim L^{\nu_{\|}/\nu_{\perp}}$.

In Monte Carlo simulations, 
a monomer is attempted to adsorb at a randomly chosen site with probability $(1-s)p$
and a dimer with probability $(1-s)(1-p)$. With probability $s$, a
randomly chosen nearest neighbor pair of adsorbed $B$'s (if there is any) 
is desorbed from the lattice.
We choose the dimer desorption probability $s=0.5$ and run stationary
Monte Carlo simulations starting with an empty lattice with size
$L=2^5$ up to $2^{11}$. The system reaches a 
quasisteady state first and stays for a reasonably long time before
finally entering into an absorbing state. 
We measure the kink density in the quasisteady state and average 
over many survived samples.
The number of samples
varies from $2\times 10^5$ for $L=2^5$ to $2 \times 10^3$ for
$L=2^{11}$. The number of time steps ranges from $10^3$ to $2\times
10^5$.

From Eq.\ (\ref{rhoL}), we expect the ratio of the critical
kink densities for two successive system sizes
$\rho(L/2)/\rho(L)=2^{\beta/\nu_\perp}$, ignoring corrections to
scaling. This ratio converges to unity in the active phase ($p<p_c$) and to 2 
in the absorbing phase ($p>p_c$) in the limit $L\rightarrow \infty$. 
We plot the logarithm of
this ratio divided by $\ln 2$ as a function of $p$ for $L=2^6,\cdots,2^{11}$
in Fig.\ 1. The crossing points between lines
for two successive sizes converge slowly due to strong corrections to
scaling. In the limit $L\rightarrow\infty$, we estimate the crossing
points converge to the point at $p_{c}=0.425(4)$ and
$\beta/\nu_{\perp} = 0.49(3)$. The value of $\beta/\nu_{\perp}$ agrees well
with the standard DI value 0.50.

In Fig.\ 2, we show the time dependence of the
critical kink densities $\rho(p_{c},t)$ for various system sizes
with $p_c=0.425$.
From the slope of $\rho(p_{c},t)$ we estimate 
$\beta/\nu_{\|}=0.275(5)$.
Insets show the size dependence of the relaxation time $\tau(p_{c},L)$   
and the steady-state kink density $\rho(p_{c},L)$ at criticality.
We estimate $\nu_{\|}/\nu_{\bot}=1.74(4)$ and 
$\beta/\nu_{\bot}=0.494(6)$, respectively. All of these results are 
in excellent agreement with the DI values. 

We run dynamic Monte Carlo simulations with various initial configurations
and get a more precise estimate of
the critical probability $p_{c}=0.425(1)$. 
Our estimates for the dynamic
scaling exponents are $\delta + \eta = 0.28(1)$ and 
$z=1.14(1)$~\cite{note1}, where
$\delta +\eta$ characterizes the growth of the number of kinks averaged
over survived samples and $z$ the spreading of the active 
region~\cite{Grass-Torre-79}.
These values are also in excellent agreement with the DI values.

To check the importance of the Ising symmetry among the
absorbing states, we introduce a symmetry breaking field such
that the monomer adsorption attempt at an even-numbered site is
rejected with probability $h$~\cite{Park-95d}. For finite $h$,
the O group of absorbing states 
is probabilistically preferable to the E group.
We set $h=0.1$ and run stationary Monte 
Carlo simulations for lattice sizes $L=2^5$ up to $2^9$. 
In Fig.\ 3, we plot
$\ln[\rho(L/2)/\rho(L)]/\ln2$ versus $p$, from which we 
estimate $p_{c}=0.304(2)$ and $\beta/\nu_{\perp}=0.24(1)$. 
The value of $\beta/\nu_{\perp}$ is clearly different from the
DI value but agrees well with the standard DP value 0.2524(5).
More detailed study including dynamic Monte Carlo simulations
confirms that the systems with finite $h$ belong to
the DP universality class~\cite{note1}.

Similar to the case of the ordinary IMD model, the symmetry-breaking field
makes the system behave like having only one (preferred) group
of absorbing states~\cite{Park-98}. 
Evolutions of the critical interfaces (active region)
$(a)$ for the symmetric case ($h=0$) and $(b)$ 
for the asymmetric case ($h=0.1$) are shown in Fig.\ 4.
In the symmetric case, the interfaces between the O and E group
of absorbing states diffuse until they meet and form a loop 
to disappear, which is the essential characteristic of
the DI universality class. In the asymmetric case, 
the absorbing region of the unpreferred (E) group
quickly vanishes and the interfaces between the
different groups become irrelvant. The interfaces
inside the preferred (O) group, which can disappear by themselves 
without forming loops, become dominant 
and force the system into the DP universality class.

When the desorption process of a nearest neighbor $BB$ pair
is forbidden $(s=0)$, the system can find many more absorbing states
with $BB$ pairs, e.g., ($\cdots B0A0BB0A0A\cdots$), 
in addition to the two groups of the absorbing states for $s\neq 0$.
These new extra absorbing states are generically mixtures 
of the O and E group of the absorbing states.
The O and E groups are now connected dynamically via new mixture-type
absorbing states. Consider a configuration with an interface between
two absorbing states in the different groups, 
($\cdots B0A0\underline{00}0A0A\cdots$),
where the interface is placed in two central vacancies $\underline{00}$.
With nonzero $s$, this configuration never evolves into an absorbing state.
However, in the case of $s=0$, it can evolve into a mixture-type absorbing state
by adsorbing a dimer $BB$ in the center. Actually, any interface can disappear
by itself in a finite amount of time, so there is no infinite dynamic barrier 
between absorbing states. Therefore the evolution of the interfaces
resembles the asymmetric case in Fig.\ 4.

Absorbing states for $s=0$ no longer possess the clear-cut
global symmetry which drives the system into the DI class. So we expect the
system falls into the DP class like the other IMA models without
an extra symmetry. We run dynamic Monte Carlo simulations 
starting with a lattice occupied by monomers at alternating sites except
at the central vacant site, ($\cdots A0A0\underline{0}0A0A\cdots$), 
where $\underline 0$ represents a defect.
In Fig.\ 5, we plot three effective exponents
against time; $\delta (t)$, $\eta (t)$, and $z(t)$~\cite{Grass-Torre-79}.
Off criticality, these plots show some curvatures. The values of
the dynamic scaling exponents can be extracted by taking the asymptotic
values of the effective exponents at criticality.
From Fig.\ 5, we estimate $p_{c}=0.105(1)$, $\delta=0.02(1)$,
$\eta=0.48(5)$, and $z=1.33(5)$. The values of $\delta+\eta$ and $z$ are in 
good agreement with those of the DP values~\cite{note2}. 
Introduction of the symmetry breaking field $h$ only changes the location of 
$p_c$. Stationary Monte Carlo simulations also confirm our results~\cite{note1}.

In summary, we found the first IMA model that does not belong to the DP
class, but belong to the DI class. This can be achieved by imposing a
global Ising symmetry on the absorbing states, i.e., making the two 
equivalent group of IMA states that are dynamically separated infinitely far apart. 
When the symmetry between these groups is broken, 
one group of absorbing states becomes completely obsolete and
the evolution morphology changes from a loop-like to a tree-like structure,
which ensures the system in the conventional DP class. We also found
that the system goes back to the DP class if the mixture-type absorbing
states between the two groups are added. These extra absorbing states
connect the two separated groups dynamically and make the loop-forming
process of the interfaces irrelvant.
The absorbing states in all other previously studied
IMA models are 
dynamically connected in the sense as mentioned above. 
This may explain why those models belong to the DP class.

HP wishes to thank M. den Nijs for his hospitality during his
stay at the University of Washington where this work was completed.
This work was supported by Research Fund provided by Korea
Research Foundation, Support for Faculty Research Abroad (1997).



\begin{figure}
\centering
\includegraphics[height=8cm,angle=270]{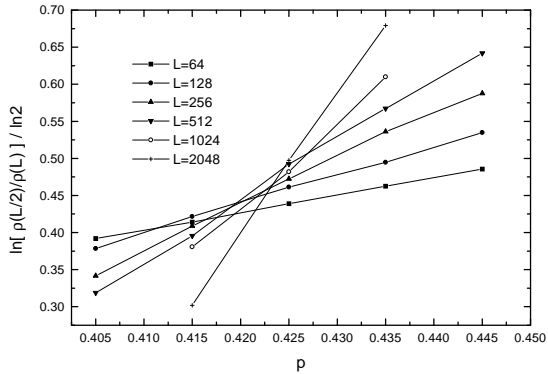}
\caption{Plots of $\ln[\rho(L/2)/\rho(L)]/\ln2$ versus
$p$ for the symmetric case.}
\end{figure}
\begin{figure}
\centering
\includegraphics[height=8cm,angle=270]{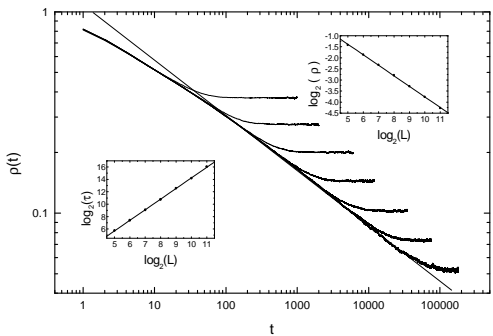}
\caption{The time dependence of the kink density at $p_c=0.425$.
The straight line is of slope 0.275 $(=\beta/\nu_\|)$.
Insets show the size dependence of the relaxation time $\tau$
and the steady-state kink density $\rho$ at criticality.
The solid lines are of slope 1.74 $(=\nu_\| /\nu_\perp)$
and $-0.494$ $(=-\beta/\nu_\perp)$.
} 
\end{figure}

\begin{figure}
\centering
\includegraphics[height=8cm,angle=270]{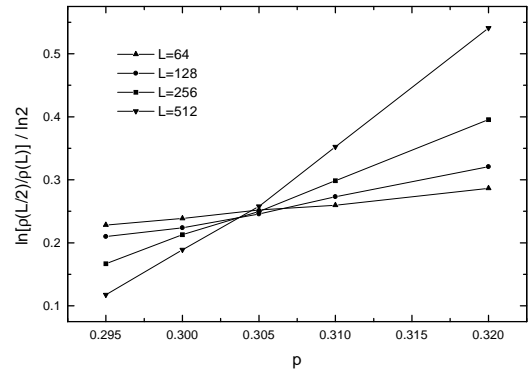}
\caption{Plots of $\ln[\rho(L/2)/\rho(L)]/\ln2$ versus
$p$ for the asymmetric case with $h=0.1$.}
\end{figure}

\begin{figure}
\centering
\includegraphics[width=8cm]{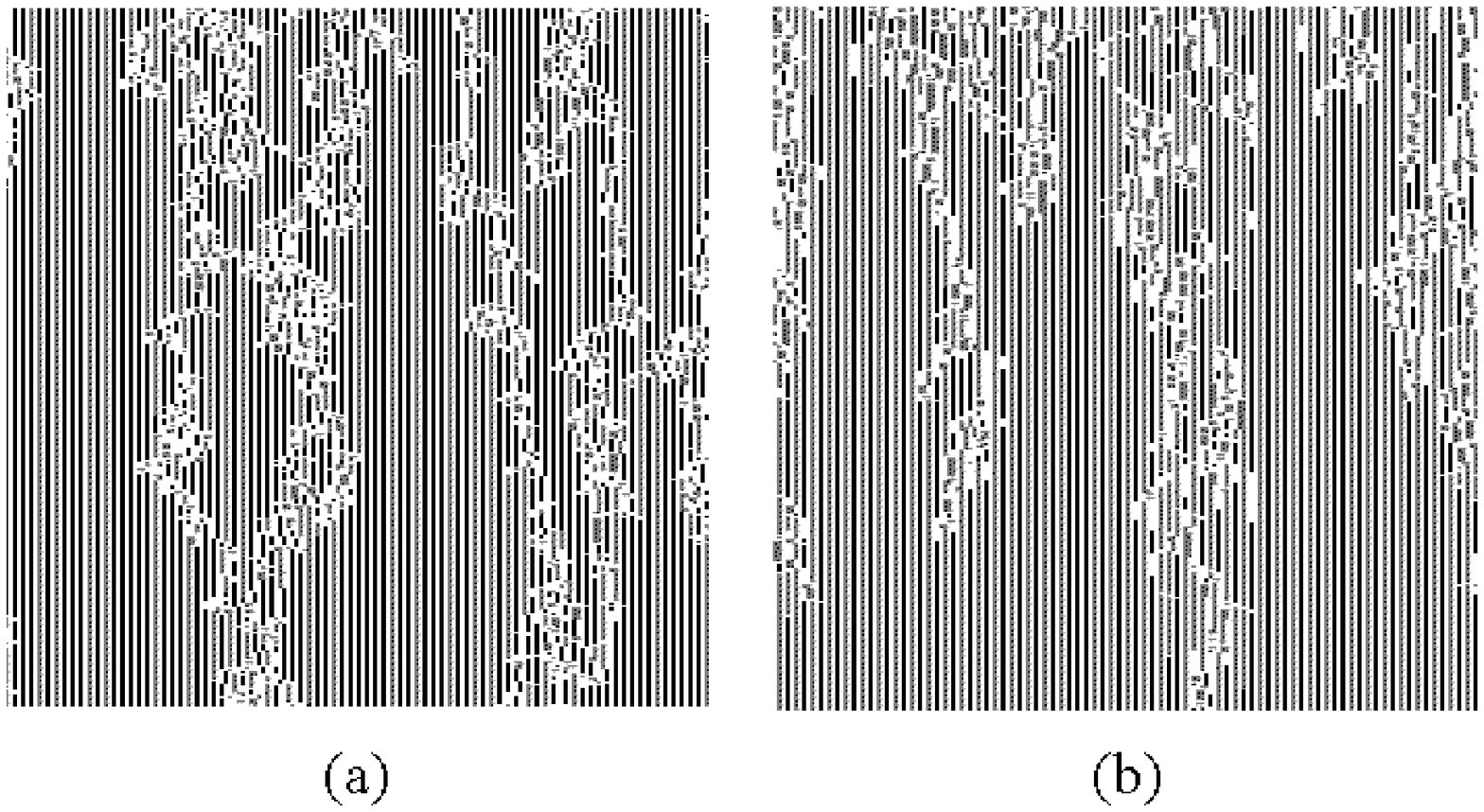}
\caption{
Evolutions of the critical interfaces for 
$(a)$ the symmetric case and $(b)$ the asymmetric case. 
Monomers $(A)$ are represented by
black, dimer particles ($B$) by grey, and vacancies by white pixels.}
\end{figure}

\end{multicols}
\widetext
\begin{figure}
\centering
\includegraphics[height=16cm,angle=270]{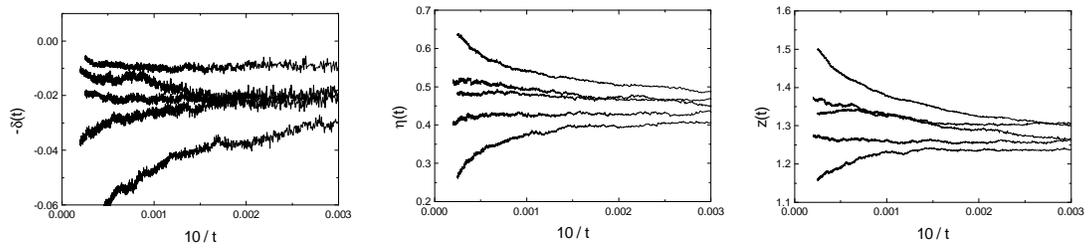}
\caption{Plots of the effective exponents against $10/t$
for $s=0$ and $h=0$. Five curves from top to botton in each panel
correspond to $p=0.100$, 0.104, 0.105. 0.107, and 0.110.}
\end{figure}


\begin{references}
\bibitem{note1} Detailed numerical results will be published elsewhere.
\bibitem{note2} The values of the exponents, $\delta$ and $\eta$, depend
on initial configurations, but their sum is universal~\cite{Mendes-94}.
\end{references}
\end{document}